\documentclass[twocolumn,superscriptaddress,preprintnumbers,amsmath,10pt,aps,prl,nobibnotes]{revtex4} 
\usepackage{amsmath}
\usepackage{amsfonts}
\usepackage{bm}
\usepackage{xcolor}
\usepackage{verbatim}
\usepackage{graphicx}
\usepackage{siunitx}
\usepackage{upgreek}

\begin{document}

\title{Time of flight 3D imaging through multimode optical fibres} 

\author{Daan~Stellinga}
\thanks{These authors contributed equally to this work}
\affiliation{School of Physics and Astronomy, University of Glasgow, G12 8QQ, UK.}
\author{David~B.~Phillips}
\thanks{These authors contributed equally to this work}
\affiliation{School of Physics and Astronomy, University of Exeter, Exeter, EX4 4QL. UK.}
\author{Simon~Peter~Mekhail}
\thanks{These authors contributed equally to this work}
\affiliation{School of Physics and Astronomy, University of Glasgow, G12 8QQ, UK.}
\author{Adam~Selyem}
\affiliation{Fraunhofer Centre for Applied Photonics, G1 1RD, Glasgow, UK.}
\author{Sergey~Turtaev}
\affiliation{Leibniz Institute of Photonic Technology, Albert-Einstein-Straße 9, 07745 Jena, Germany.}
\author{Tom\'{a}\v{s}~\v{C}i\v{z}m\'{a}r}
\affiliation{Leibniz Institute of Photonic Technology, Albert-Einstein-Straße 9, 07745 Jena, Germany.}
\affiliation{Institute of Scientific Instruments of the CAS, Královopolská 147, 612 64 Brno, Czech Republic.}
\author{Miles~J.~Padgett}
\email{miles.padgett@glasgow.ac.uk.}
\affiliation{School of Physics and Astronomy, University of Glasgow, G12 8QQ, UK.}

\begin{abstract}
{\bf Time-of-flight (ToF) 3D imaging has a wealth of applications, from industrial inspection to movement tracking and gesture recognition.
Depth information is recovered by measuring the round-trip flight time of laser pulses, which usually requires projection and collection optics with diameters of several centimetres. In this work we shrink this requirement by two orders of magnitude, and demonstrate near video-rate 3D imaging through multimode optical fibres (MMFs) - the width of a strand of human hair. Unlike conventional imaging systems, MMFs exhibit exceptionally complex light transport resembling that of a highly scattering medium. To overcome this complication, we implement high-speed aberration correction using wavefront shaping synchronised with a pulsed laser source, enabling random-access scanning of the scene  at a rate of $\sim$23,000 
points per second. Using non-ballistic light we image moving objects several metres beyond the end of a $\sim$40\,cm long MMF of 50\,$\mu$m core diameter, with millimetric depth resolution, at frame-rates of $\sim$\,5Hz. Our work extends far-field depth resolving capabilities to ultra-thin micro-endoscopes, and will have a broad range of applications to clinical and remote inspection scenarios.}
\end{abstract}

\maketitle

\noindent{\bf Introduction}.
Multimode optical fibres (MMFs) represent an extremely efficient method of transporting light with a high spatial information density. They can support the propagation of thousands of spatial modes - i.e. optical field patterns which act as independent information channels - within a cross-sectional area similar to that of a human hair. These features have led to much interest in the deployment of MMFs as micro-endoscopes, enabling high-resolution imaging at the tip of a needle~\cite{Choi,ohayon2018minimally,turtaev2018high,vasquez2018subcellular,amitonova2020endo}. In this work we investigate the feasibility of enhancing MMF-based imaging to include depth information. Extending 3D imaging capabilities through ultra-thin MMFs promises an array of new applications, including the 3D inspection of the internal chambers of objects that are difficult to open, such as jet engines or nuclear reactors, and the 3D visualisation of hollow viscous organs, which could help surgeons navigate inside the body during operations.

However, the compact form of MMFs comes at a cost: monochromatic optical signals are subject to modal dispersion, as the phase velocity of light propagating through a MMF depends upon its spatial mode. Therefore input coherent light patterns are typically unrecognisably scrambled into speckle patterns at the output facet, formed entirely from non-ballistic light that has scattered multiple times from the core-cladding interface~\cite{goodman1976some,ploschner2015seeing,Velsink:21}. Fortunately, as long as a MMF remains in a fixed configuration, the scrambling process is deterministic and unchanging in nature, and so a given input field will always produce the same output field. This means that the way a static MMF scrambles light can be represented by a linear matrix operator, known as a transmission matrix (TM), which maps any possible input field to the resulting output~\cite{popoff2010measuring,Cizmar:11,vcivzmar2012exploiting,popoff2010image,Papadopoulos:12,DiLeonardo:11,li2021compressively}.

Measurement of the TM enables calculation of how an input field should be pre-shaped to generate a desired output, for example a spot focused to a particular location. This method is known as wavefront shaping~\cite{Vellekoop:07,mosk2012controlling}, and thus by illuminating the proximal facet with a sequence of carefully prepared input light fields, a focused spot can be raster scanned across the distal facet of a MMF. Scanning-based imaging can then be achieved by recording the total intensity of light returning through the fibre and correlating this with the position of the focus~\cite{gusachenko2017raman,tragaardh2019label}.
\begin{figure*}[t!]
\centering
\includegraphics[width=0.8\textwidth]{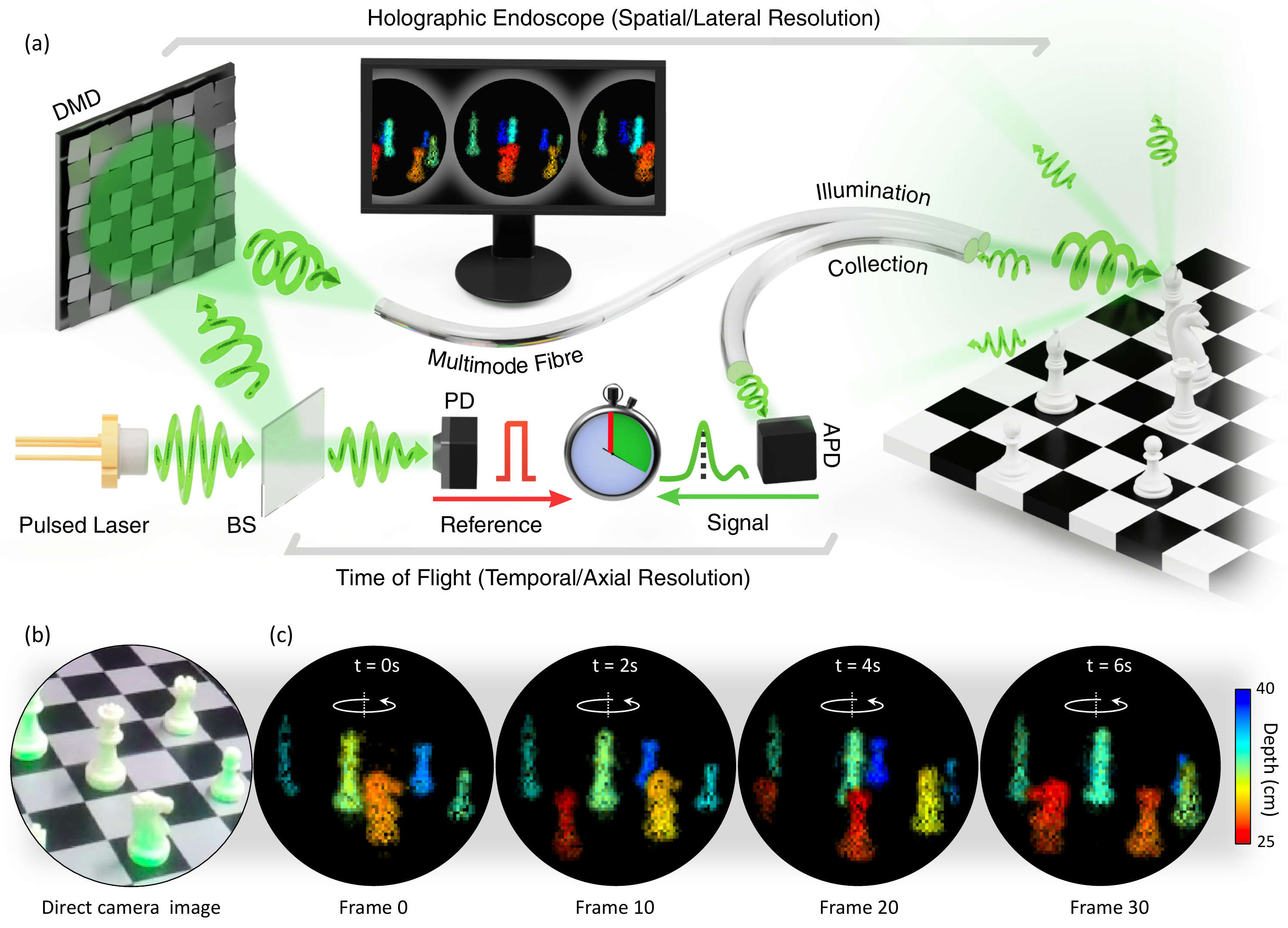}
\caption{\textbf{Endoscopic LiDAR.} (a) A schematic of the experimental set-up. (b) A snapshot of the true scene being recorded. (c) Typical depth-resolved images obtained with the our system. Each frame is captured in 200\,ms. The frames show the pieces on a revolving chess board located at a depth of $\sim$30\,cm, recorded at a frame rate of 5Hz. The dark spots in images are due to singularities in the speckle reference used to measure the TM.}
\end{figure*}

Recently wavefront shaping through MMFs has been employed for in-vivo imaging of neurons deep inside the brains of mice~\cite{ohayon2018minimally,turtaev2018high,vasquez2018subcellular} - an endeavour very challenging to achieve in any other such minimally invasive way. Imaging of objects some distance from the distal facet of a MMF is also possible~\cite{vcivzmar2012exploiting,leite2020observing}. However, this is more demanding, since the level of return signal falls off rapidly in proportion to the square of the objects’s distance.

Here we augment MMF micro-endoscopy with time-of-flight (ToF) LiDAR (Light Detection And Ranging) techniques, to provide depth information alongside 2D reflectance images~\cite{schwarz2010mapping,edgar17,tobin2019three}. ToF techniques recover depth by measuring the round-trip flight time of a laser pulse reflecting from the scene. To achieve this we implement high-speed wavefront shaping of a sub-ns pulsed laser source. We measure the TM linking the proximal end of a MMF to the far-field of the distal facet, and calculate the input fields required to raster scan a focused spot across the far-field scene. A second MMF collects the back-scattered light, which is coupled to a fast photo-detector, enabling measurement of both the reflectivity and the time-of-flight at each spot location. The two MMFs therefore act as an endoscopic LiDAR system with highly compact projection optics of $\sim$600 microns in diameter - roughly two orders of magnitude smaller than conventional LiDAR based topographic imaging systems. Our endoscope can deliver depth resolved images of macroscopic scenes up to 2.5\,m away from the fibre facets with an imaging frame-rate up to 5\,Hz. Figure 1 shows a schematic of our setup and some example images, each recorded in 200\,ms - here identifying the relative depths of chess pieces on a revolving chess board positioned $\sim$30\,cm from the distal fibre facet.\\

\noindent{\bf Results}.
Extending wavefront shaping through MMFs from continuous-wave to pulsed illumination presents an additional complication: the potential for temporal pulse distortion due to chromatic or spatial mode dispersion. In our system the latter form of dispersion dominates. Spatial mode dispersion can be understood by considering that an incident pulse will excite different spatial modes supported by the fibre, which travel at slightly different velocities~\cite{carpenter2017comparison,Velsink:21b,xiong2019long}. A MMF acts as a multi-path interferometer with many different arms - about 1000 in our case. Once the optical path difference between these arms, $\Delta_{\rm{OPL}}$, is greater than the coherence length of the pulse, $\ell_{\rm{p}}$, light in these respective spatial modes no longer interferes, and the visibility of the output interference pattern reduces. In the extreme this leads to severe temporal pulse distortion - whereby a single input pulse fragments into separate pulses propagating in different spatial modes that exit the fibre in sequence and so are unable to interfere with one another~\cite{johnson2019light}. In this case wavefront shaping, which relies on controlling the interference of different spatial modes at the output, is not possible without spatio-temporal compensation~\cite{mounaix2016spatiotemporal,mounaix2020time}. To avoid the added complexity of spatio-temporal beam shaping, we must ensure that $\ell_{\rm{p}} >> \Delta_{\rm{OPL}}$. This in turn places constraints on the temporal length of the pulse $\tau_{\rm{p}}$ that can be used with a given geometry of the fibre, leading to (see Supplementary for derivation):
\begin{equation}\label{Eqn:time}
    \tau_{\rm{p}} >> \frac{\rm{NA}^2L}{cn_{\rm{c}}}.
\end{equation}
where $L$ is the (single pass) fibre length, $\rm{NA}$ is the numerical aperture of the fibre, $n_{\rm{c}}$ is the refractive index of the core, and $c$ is the speed of light in a vacuum. Here we use a fibre of $L\sim0.4$\,m, $\rm{NA} = 0.22$, and $n_{\rm{c}} = 1.45$, meaning $\tau_{\rm{p}}$ should be significantly longer than $\sim$45\,ps. To ensure minimal pulse distortion, we choose a laser with a pulse duration of $\tau_{\rm{p}}\sim$700\,ps, i.e.\ a factor of $\sim$15$\times$ longer than the limit set by Eqn.~\ref{Eqn:time}.

The number of independently resolvable features in images that can be transmitted through a MMF is proportional to the number of spatial modes, $N$, it supports per polarisation degree of freedom. $N\sim\left(\pi a\rm{NA}/\lambda\right)^{2}$, where $a$ is the radius of the fibre core, and $\lambda$ the illumination wavelength. In our case, $a=25$\,$\mu$m and the central wavelength of the pulsed source is $\lambda = 532$\,nm. Therefore $N\sim 1000$, which means it is possible to project $\sim$\,1000 non-overlapping foci within the field-of-view. This sets the lateral resolution of our system to $\sim 4N = 4000$ independently resolvable features within each image, when resolution is defined by the Raleigh criterion~\cite{mahalati2013resolution,leite2020observing}.

Before the system can be used for imaging, we must first calibrate the TM of the MMF. Using our pulsed source we measure the TM linking the field at the input side of the MMF to the far-field of the distal facet. Wavefront shaping is achieved with a high-speed digital micro-mirror device (DMD)~\cite{conkey2012high,turtaev2017comparison}. During TM measurement, the DMD is used to scan a combination of two spatial modes at the entrance facet - a fixed reference mode, and a changing probe mode, which co-propagate through the fibre. At the output side of the MMF the reference and probe modes emerge from the distal facet and propagate $\sim$20\,cm through free-space to form an interference pattern on a screen placed in the far-field of the fibre. During the calibration step we image this interference pattern with a CMOS camera, synchronised with the DMD. Using phase stepping holography we recover the complex field on the screen, providing the relation between input and output field for each probe mode - and thus build the TM column by column. Supplementary Information provides more details.

Once the TM is measured, it is used to calculate a sequence of DMD patterns that shape the wavefront of the input pulses in such a way that they are transformed into spots in the far-field of the distal fibre facet. These spots can be laterally raster-scanned over the field-of-view at a rate of up to $\sim$23,000 points per second when imaging. See Supplementary Information for more details. The MMF must remain in a fixed configuration after TM calibration, although we note that in our experiments this was achieved simply by holding each end in a standard fibre connector, without further mechanical or temperature stabilisation. In this configuration a single TM can still be effectively used for several days. To maximise the power projected into the focused spots, we use phase-only wavefront shaping, and place the DMD in the image plane of the fibre facet~\cite{leite2020observing}.

During the recording of ToF images, a second fibre is used to collect the backscattered light, placed alongside the illumination fibre. Both fibres have the same NA of 0.22, however the collection fibre has a larger core diameter of 500\,$\mu$m to increase the collection efficiency and thereby the working distance of the endoscope. The alternative approach of conducting both illumination and detection through a single fibre is challenging due to reflections and scattering at both fibre facets. The small size of the collection aperture means that only in the order of 1 in $10^{10}$ backscattered photons will be reflected back into the fibre from an object 1\,m away. Even an antireflection coated fibre would still reflect upwards of 0.1\% of transmitted light at both facets, overwhelming the return signal from the scene itself. Future systems may overcome this limitation, and we envisage a double clad fibre with illumination delivered through a central fibre and collection through a larger diameter outer fibre may offer a compact solution.   
\begin{figure*}[t!]
\centering
\includegraphics[width=0.8\textwidth]{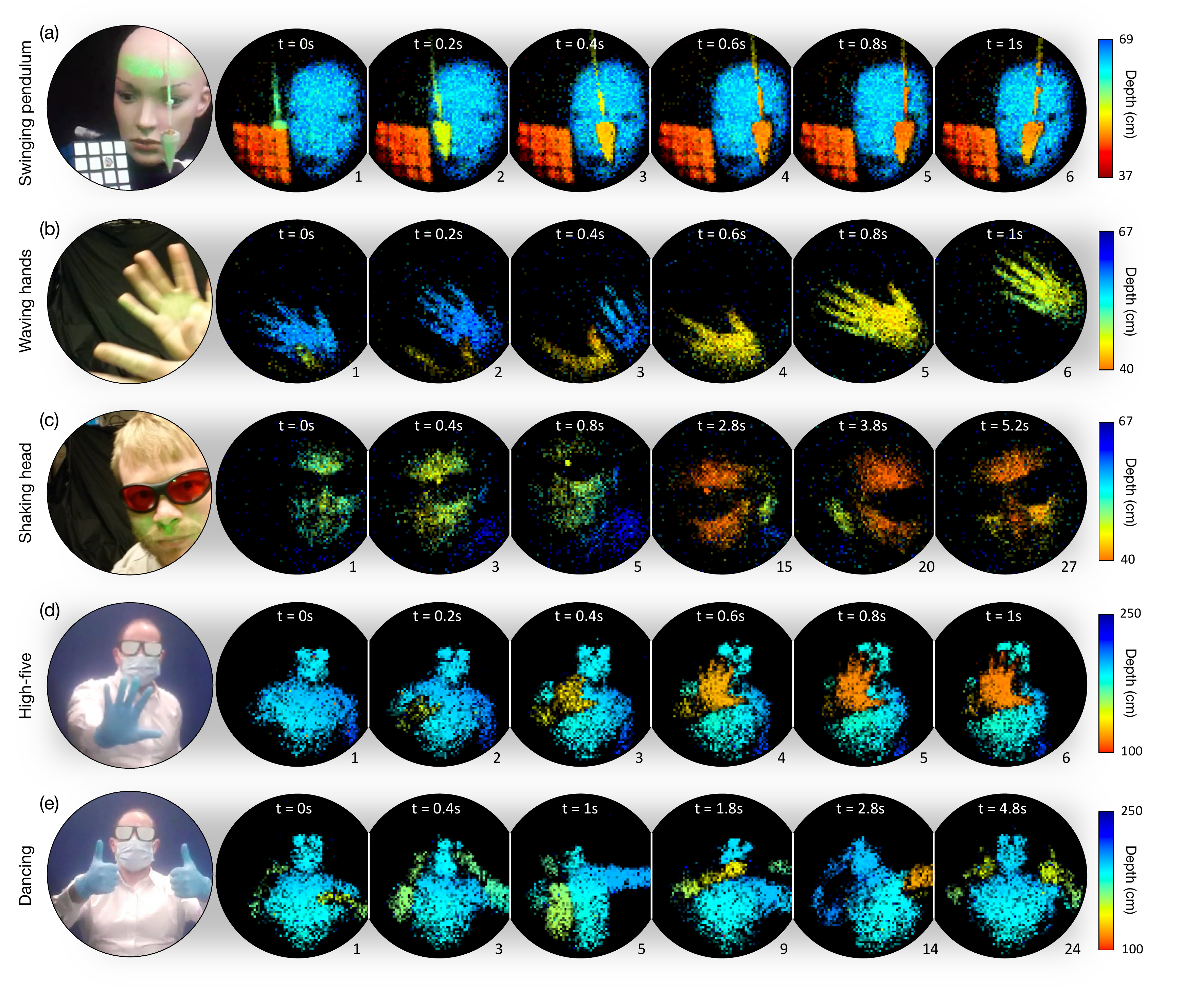}
\caption{\textbf{Snapshots from depth resolved movies at progressively increasing scene depths.} Left hand column shows a direct camera image of each scene. (a) Consecutive frames of a pendulum swinging adjacent to a mannequin head and a Rubik's cube in close range to the distal fibre facet (0-26cm). (b,c) Author Daan Stellinga waving his hands and shaking his head at a range of 40-67cm. (d,e) Author Simon Peter Mekhail giving a `high-five' and dancing at a range of 1-2.5\,m. In these images, scene depth is encoded in the colour channels, and scene reflectivity is encoded in the transparency channel - thus regions of the scene with low reflectivity, and consequently a poorly estimated depth, are displayed with low brightness. The relative time is given at the top of each frame. The frame number is given at the bottom-right of each frame.}
\end{figure*}

Light that has scattered from the scene enters the collection fibre, and is then coupled directly to an avalanche photodiode (APD). The APD signal is fed into a high speed digitiser triggered by a reference signal from a photodiode detecting the time that each input pulse enters the illumination fibre. The digitiser samples at 2.5\,GS/s, equating to time bins of 400\,ps. This is sufficient to resolve the $\sim$700\,ps pulses, enabling measurement of the flight time and peak intensity of the reflected pulse, and thus calculation of both the depth and the reflectivity of the pixel being scanned. The depth of each pixel, $D$, is a function of both the measured time of flight of the pulse, $\delta t$, and the angular coordinate of the pixel with respect to the optical axis of the fibre $\theta$:
\begin{equation}\label{Eq:flightDist}
    D\left(\delta t,\theta\right) = \frac{1}{2}c\delta t\cos\theta - n_{\rm{c}}L\cos\theta\left(1-\frac{\sin^2\theta}{n_{\rm{c}}^{2}}\right)^{-\frac{1}{2}}.
\end{equation}
Equation~\ref{Eq:flightDist} accounts for the different optical path lengths travelled by pulses propagating through, and emanating from, the fibre at different angles. See supplementary Information for a derivation. We note that in our case, the second term on the right-hand-side is negligible for the short and relatively low $\rm{NA}$ of fibres used, and so was omitted in data processing.

The repetition rate of the laser is selected to be just below the maximum refresh rate of the DMD (22.7\,kHz), giving exactly one laser pulse per image pixel. We measure $\sim$4200 pixels per image frame, thus moderately oversampling  the scene to improve the signal-to-noise ratio (SNR) of the system. This equates to a near video frame-rate of 5 frames per second.

Figure 2 shows snapshots from videos of several dynamic scenes. All frames are recorded in 200\,ms. Figure 2a shows consecutive frames of a pendulum (of $\sim$1\,m in length), swinging in between two static objects, positioned $\sim$0.4-0.6\,m from the fibre facets. At this range the depth map tracks the 3D motion of both the pendulum bob and its thread. A small amount of rolling shutter effect is visible, causing the bob to appear to point in the direction of movement, due to the raster scan reaching the tip slightly later in the swing than the base. Figures 2b-e show scenes with more dynamic motion: some of the authors moving around at depths of $\sim$0.4-0.7\,m (2b,c) and further away at $\sim$1-2.5\,m (2d,e). The full videos are available in supplementary information. These depict the 3D nature of the scene, and independently show scene reflectivity and depth estimation.\\

\noindent{\bf Discussion}.
To quantitatively assess the imaging performance of our prototype system, we measure the depth error,
angular resolution, spot contrast ratio and signal-to-noise-ratio across the field-of-view.

The depth precision is dependent upon how well the round-trip flight time of the laser pulse, $\delta t$, can be estimated (see Eqn.~\ref{Eq:flightDist}). Under the assumption that each pulse reflects from a single interface and the returning pulse shape is not distorted, then $\delta t$ is not limited by pulse duration $\tau_{\rm{p}}$, and can be estimated beyond the temporal sampling resolution (of $t_{\rm{s}} = 400$\,ps in our case). This can be achieved, for example, by digitally upsampling the recorded histograms and finding the time of the peak in the cross-correlation between the histogram and the known pulse shape. Ultimately this protocol is limited by the point at which further upsampling becomes dominated by measurement noise. Here we use a simple algorithm capable of real-time operation: we upsample by a factor of $N_{\rm{s}}=32$ using sinc interpolation and register the time of peak intensity. This provides a nominal depth precision of $\sim t_{\rm{s}}c/(2N_{\rm{s}})\sim$\,2\,mm, although we note the true value is scene specific and dependent on the level of measurement noise, which grows with the distance to the scene. In addition, objects with smooth surfaces yield higher precision depth estimation by minimising distortion of the return pulse, or its fragmentation into several pulses arriving at different times. Figure 3a shows a plot of the measured depth error as a function of angular spot position within the field-of-view, showing both accuracy and precision. Supplementary information gives more details about this measurement.
\begin{figure}[t!]
\centering
\includegraphics[width=0.45\textwidth]{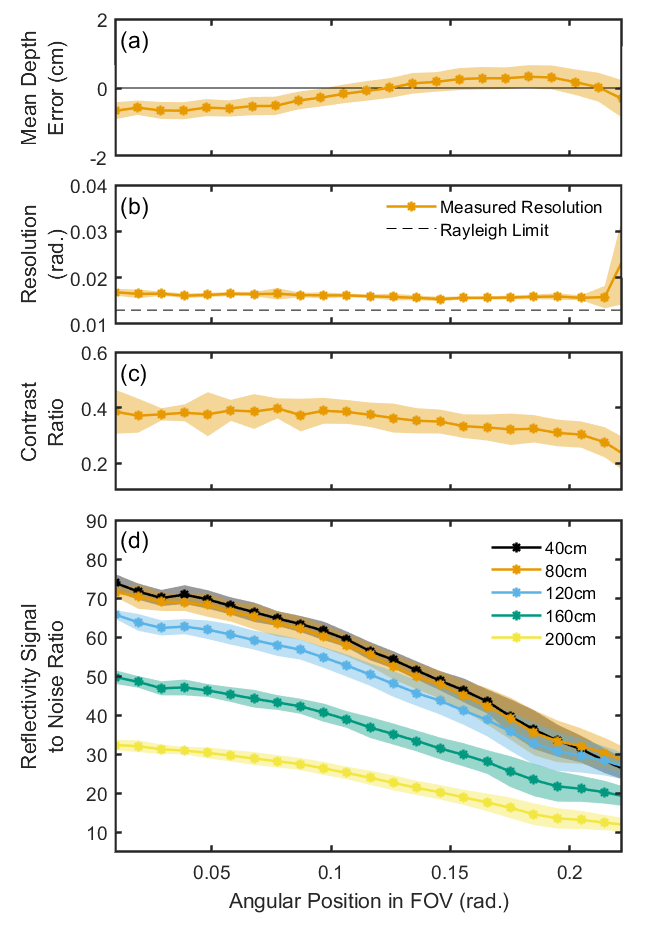}
\caption{\textbf{System characterisation as a function of radial position in field-of-view} (a) Mean error in depth estimation. The points indicate the depth accuracy, the shaded area indicates the depth precision via the standard deviation calculated from 100 repeat measurements. (b) Angular resolution estimated by fitting Airy discs to the projected spot profiles. (c) Spot contrast ratio. (d) Signal-to-noise ratio as a function of object depth.}
\end{figure}

In the far-field of the illumination fibre facet, the lateral resolution is well approximated by Fraunhofer diffraction from a circular aperture the size of the fibre core. Therefore the radius $r$ of a spot at a depth $l$ from the distal fibre facet is given by $r\sim 0.61 l\lambda/a$. Since the number of resolvable features is fixed by the number of spatial modes supported by the MMF, both the lateral resolution and the diameter of the field-of-view grow linearly in proportion to $l$ as the distance to the scene increases. The angular resolution limit, $\theta_r$ is constant with distance and given by $\theta_r\sim r/l\sim 0.61\lambda/a$. Figure 3b shows the measured angular resolution is uniform and close to the Fraunhofer diffraction limit across most of the field-of-view, only increasing in size at the edges when projected from the distal facet beyond an angle of $\sim$0.2\,rads., which is approaching the $\rm{NA}$ of the fibre (0.22\,rads.).

Figure 3c shows how the spot contrast ratio, defined as the ratio of power in the spot to the total projected power, varies across the field-of-view. The contrast ratio is typically 0.4, and reduces slightly towards the edge of the field-of-view. In this case the spots are projected using phase-only wavefront shaping, with the DMD placed in a plane conjugate to the MMF input facet. This configuration maximises the total projected power at the expense of a reduction in contrast ratio. Additionally, contrast ratio is also reduced by any small drifts in the setup after the TM is measured, or heating of the fibre itself.

Figure 3d shows how the signal-to-noise ratio (SNR) of our system depends upon both the angular position and the depth of the object. Here SNR is defined as the ratio between the temporal histogram peak and the standard deviation of the surrounding data points. At all depths, SNR is highest in the centre of the field-of-view and reduces towards the edge. This is because spots projected at higher angles from the distal facet are formed from rays propagating closer to the critical angle of total internal reflection, and consequently suffer the greater levels of power leakage into the cladding. The collection efficiency, and thus SNR, reduces with increasing object depth as expected.

There are several ways the image quality can be enhanced in our current prototype system if necessary. Although $\tau_{\rm{p}}$ does not directly limit the depth precision, shorter pulses - providing they can be properly sampled - will yield a higher depth precision. SNR can be improved by reducing the frame-rate to average more laser pulses per pixel. Blind spots in the images also appear at the locations of the vortex singularities in the reference speckle pattern. These can be removed by conducting the TM calibration with an external reference beam that is guided around the fibre, or by integrating multiple speckle references which are unlikely to have blind spots at coincident locations. Real-time images shown in Figs 1 and 2 are reconstructed by assuming that the scene was illuminated with an ideally focused spot. However, drift in the system, the use of an internal reference beam, and the phase only nature of the wavefront shaping all contribute to reduce the contrast of the focused spot above a speckled background which appears across the entire field-of-view. If the actual projected patterns are measured during the calibration phase and do not change, then a more sophisticated reconstruction algorithm can be used to incorporate this information. We term this hybrid method TM guided computational imaging: SNR is maximised by using the TM to concentrate power into a focused spot, and regularised matrix inversion is then used to account for residual background speckle. Figure 4 depicts a series of views of a static 3D image of a mannequin head reconstructed from data recorded over 2\,s using TM guided computational imaging. The depth error has been reduced to an extent that enables the contours of the face to be revealed. Supplementary information gives more detail of the TM calibration and reconstruction method used in this case.

We note that Eqn.~\ref{Eqn:time} provides a bound on the minimum pulse duration, as a function of fibre length, to achieve arbitrary wavefront shaping at the output without temporal distortion. However for the special case of the generation of focused points in the far-field of the distal facet of a MMF, pulses of significantly shorter duration than this bound may be used without distortion. This is because, due to the approximate cylindrical symmetry of a MMF, each far-field point only requires the excitation of spatial modes with very similar phase velocities, i.e.\ spatial modes with large differences in phase velocity are never excited simultaneously. Therefore far-field endoscopic LiDAR could potentially be achieved through MMFs of greater length, or with a shorter pulse duration, than implied by Eqn.~\ref{Eqn:time}.

Finally, we consider our work in the context of alternative 3D imaging approaches. There are an emerging class of clinical micro-endoscopes capable of recovering high-resolution depth images within tissue using optical coherence tomography (OCT)~\cite{fujimoto1995optical,tearney1997vivo,gora2017endoscopic,li2020ultrathin}. These endoscopes typically deliver images with an axial resolution of $\sim10\,\mu$m, and lateral resolution of $\sim30\,\mu$m, and range from a few millimetres down to $\sim500\,\mu$m in diameter. Light is delivered through a single mode fibre, and many designs feature bespoke micro-optics at the distal facet giving a side-view of the sample. Images may be constructed pixel-by-pixel by mechanically rotating and retracting the endoscope in a spiral motion, enabling 3D tomography of the tissue surrounding the endoscope shaft. Rather than imaging the cross-section of vessels and similar, our system is designed to operate in a different regime - namely to image more distant objects and surfaces, in the far-field of the fibre output, at near video frame-rates. As such our system has a greater depth range and yet lower axial resolution, which in our case is constrained by the shortest coherence length of the light it is possible to transmit through MMFs without spatio-temporal compensation (see Eqn.~\ref{Eqn:time}). The use of a MMF that supports many spatial modes also enables high-speed DMD-based point scanning rather than mechanical scanning, allowing for high frame-rate operation, while sacrificing flexibility of the endoscope itself.
\begin{figure}[t!]
\centering
\includegraphics[width=0.45\textwidth]{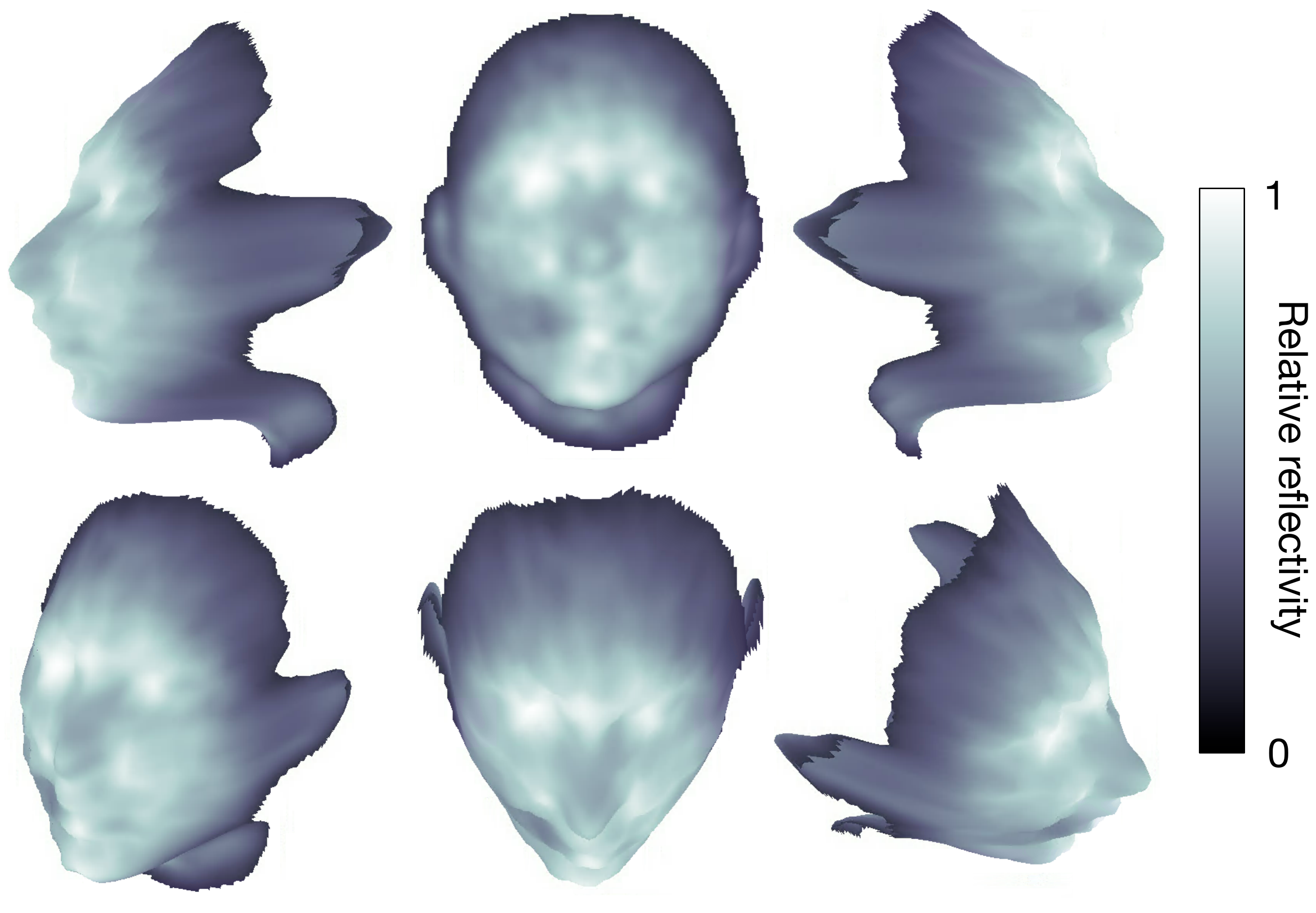}
\caption{\textbf{3D imaging with enhanced fidelity}. A series of views of a mannequin head located at a range of 30\,cm. In this case the data was recorded in 2\,s.}
\end{figure}

There are also several methods to image in 3D through scattering media under development. Time gated LiDAR is a well-established technique to see through obscurants by filtering out scattered light (by photon arrival time) to recover a signal dominated by ballistic photons that have travelled straight through the occluding object~\cite{wang1991ballistic}. More recently, non line-of-sight 3D imaging around corners has been demonstrated, relying on streak or single photon avalanche diode (SPAD) cameras to overcome signal mixing due to several diffuse reflections from opaque walls\cite{velten2012recovering,gariepy2016detection}. Our approach is fundamentally different to these methods. Our measurements are formed purely from light that has forward scattered many times through a MMF, with no ballistic component remaining. We extract image data from this non-ballistic light by pre-characterisation of a TM that enables the spatial control of the incident field to guide it through the MMF to its target. Our work also complements recent demonstrations of 3D imaging through thick randomly scattering media, in which diffuse scattering effects are inverted via ultra-fast pulse detection~\cite{lindell2020three}, and optical sectioning through larger form factor multicore fibres using in-situ distal holography~\cite{badt2021label}.\\

\noindent{\bf Conclusion}.
In summary, we have demonstrated a MMF-based micro-endoscope capable of near video-rate 3D far-field imaging. Our prototype system delivers images through a 40\,cm MMF at 5\,Hz, each frame containing up to $\sim$4000 independently resolvable features, with a depth resolution of $\sim$5\,mm. We have described how the lateral resolution, depth-precision, frame-rate and fibre length are interdependent but can be tuned by adjusting the pulse duration of the laser source, and the geometry of the MMF used. Currently the MMF has to remain in a fixed position after TM calibration. However in the future our concept could be combined with emerging techniques to monitor the TM of flexible MMFs in real-time with access to only the proximal end, and update the pre-shaped light fields accordingly\cite{li2021memory,gordon2019characterizing}. Our prototype ToF-based 3D far-field imaging system brings a new imaging modality to MMF-based micro-endoscopy, with many potential applications to remote inspection and biomedical imaging in the life sciences.

\section*{Acknowledgments}
The authors thank Ivo Leite for useful discussions, and Steven Johnson and Graham Gibson for practical advice in the lab. D.S., S.P.M \& M.J.P acknowledge financial support from the Horizon 2020 project QSORT (766970) and Quantic (EP/M01326X/1). D.B.P acknowledges financial support from the Royal Academy of Engineering and the European Research Council (804626). S.T. \& T.C. acknowledge financial support from MEYS, the European Regional Development Fund (CZ.02.1.01/0.0/0.0/15\_003/0000476) and the European Research Council (724530). M.J.P. thanks the Royal Society for financial support.

\section*{Author contributions}
\textbf{Daan Stellinga:} Conceptualization, formal analysis, investigation, methodology, writing - original draft; \textbf{David B.~Phillips:} Conceptualization, formal analysis, visualisation, writing - original draft; \textbf{Simon Peter Mekhail:} Formal analysis, visualisation, investigation, methodology, writing - original draft;  \textbf{Adam Selyem:} Software; \textbf{Sergey Turtaev:} Methodology, software, visualisation; \textbf{Tomáš Čižmár:} Methodology, conceptualization, formal analysis, writing - review \& editing, funding acquisition, supervision; \textbf{Miles J. Padgett:} Methodology, conceptualization, formal analysis, writing - review \& editing, funding acquisition, supervision.

\bibliography{fibre_refs2.bib}

\clearpage
\onecolumngrid
\begin{center}
{\large {\bf Time of flight 3D imaging through multimode optical fibres\\ - Supplementary Information}}
\end{center}
\twocolumngrid

\noindent{\bf Transmission matrix acquisition}.
For fast and precise control of the fibre input light field a DMD is used in a plane conjugate to the fibre facet. A set carrier-grating is displayed on the DMD (Vialux V-7000) which directs a portion of the laser power to a desired first order diffraction peak in the focal plane of a transforming lens. An iris in the focal plane of this lens filters all light which is not in the desired diffraction order. Further to the carrier frequency, a phase correction is added to compensate for irregularities in the flatness of the DMD. Using a relay telescope with a magnification of 4, the field at the iris is transferred to the back focal plane of an objective trained on the fibre input facet.

Here we chose to use an ‘internal’ reference beam, that propagates through the fibre rather than the more common ‘external’ reference beam that propagates around the fibre. This has the advantage of simplicity and high interferometric stability, at the cost of slightly lower fidelity measurements and a small number of missing points in the resulting raster scan - at the locations of the vortex singularities in the reference speckle pattern. If necessary these missing points can be removed using additional calibration measurements recorded with several distinct references. This extra step was only applied for the longer exposure image shown in Fig.~4.

To generate the internal reference mode, 25 spatially distinct points were excited on the back focal plane of the objective. This was done by superimposing the corresponding plane waves at the DMD which would generate these points as well as the carrier grating and converting the resultant field to a binary hologram through thresholding. Special care was taken to ensure that the points selected remain within a circular bound which would not exceed the fibre’s numerical aperture. Several plane waves were used, in this case 25, to ensure that the reference field was generated with a superposition of fields with multiple different axial components, $k_{z}$, of their wave vectors. If the reference field were generated with a single dominant $k_{z}$, the output in the far field of the distal end of the fibre would have the majority of its power localised within a ring. Such a ring shaped reference field would interfere poorly with probe fields with different $k_{z}$ values.

The configuration described above efficiently couples the DMD plane to the input fibre facet, guaranteeing an optimal use of the available laser power. An alternative configuration with the DMD in a Fourier plane of the fibre facet very slightly improves the fidelity of the spots and simplifies the required reference to a single plane wave component, but at the cost of a significant reduction in power transmission, making it less useful at longer range and higher frame rates. This alternative configuration was also tested, and used only for the results shown in Fig.~2b,c  and the associated video 3.

A total of 1961 input modes are measured using phase stepping holography. These input modes are selected from a rectilinear grid within a circular bound, as dictated by the fibre’s numerical aperture, in the back focal plane of the coupling objective. To measure an input mode, the plane wave associated with the mode is added to the reference and carrier superposition at plane of the DMD and the resultant field is converted to a binary hologram. For each input mode this is repeated four times stepping the phase of the test mode forward by $\frac{\pi}{2}$ rad.\ each time and an image is captured at every step. Further to these two more images are captured to measure the background illumination with the laser diverted away from the iris, and an image of the reference beam alone to measure power drift in the laser. Changes in these values are corrected for in the generation of the TM.

To generate the transmission matrix, all input mode measurements first go through background subtraction and normalisation by the reference-only images. Each pixel in the 180 × 180 region of interest on the CMOS (Hamamatsu Orca Flash 4.0) is then recorded for the four phase steps and the inner product these points and the expected sinusoid is taken to determine the optimal phase offset and amplitude required for the tested input mode to constructively interfere at the pixel in question. Once these values are determined for all pixels the frame is vectorised and this forms the first column of the TM. Subsequent columns are formed in the same way for the subsequent input modes.\\

\begin{figure*}[t!]
\centering
\includegraphics[width=0.8\textwidth,trim=10 10 10 0, clip]{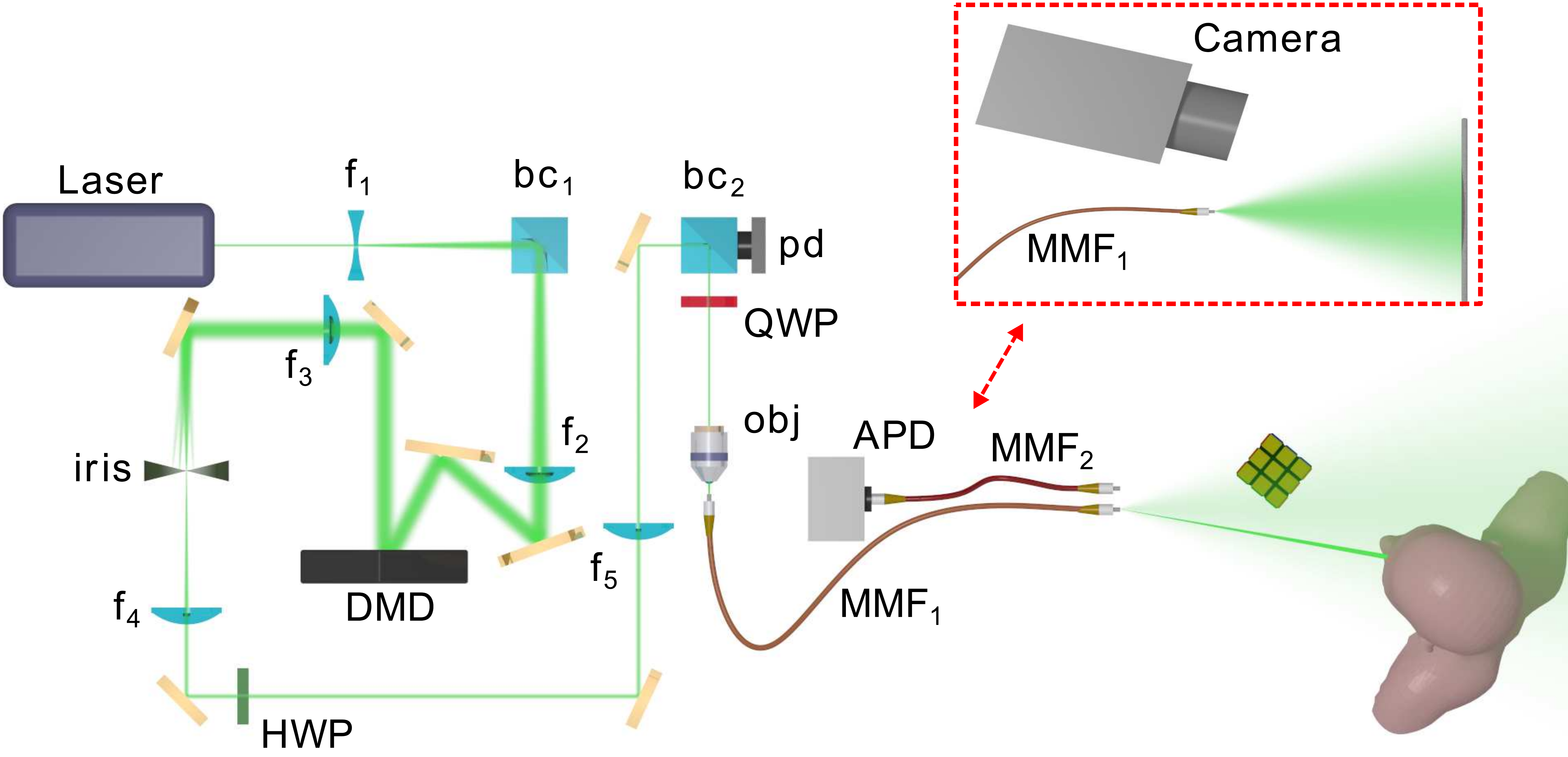}
\caption{\textbf{Experimental setup.} A passively Q-switched pulsed laser source (532\,nm, 21\,kHZ repetition rate, 700\,ps pulse width, Teem Photonics SNG-100P-1x0) is expanded and collimated onto the DMD through a combination of two lenses ($\rm{f}_{1}=-10\rm{mm}$ and $\rm{f}_{2}=300\rm{mm}$). An intermediate polarising beam cube ($\rm{bc}_{1}$) is used in reflection to ensure vertical polarisation. A single diffraction order from the DMD is filtered out through a combination of a lens ($\rm{f}_{3}=200\rm{mm}$) and an iris in a 2f configuration. A 4f telescope ($\rm{f}_{4}=100\rm{mm}$ and $\rm{f}_{5}=400\rm{mm}$) further magnifies and images the iris plane onto the back focal plane of the objective (Olympus plan-N $40\rm{x}/0.65$), with the input facet of the illuminating multimode fibre ($\rm{MMF}_{1}$ Thorlabs FG050LGA) at its working distance. The light from the distal end of this fibre freely propagates towards the scene being imaged, with a secondary multimode fibre ($\rm{MMF}_{2}$ Avantes FC-UV600) placed right next to the first collecting the backscattered light. This second fibre is directly coupled to the sensor of an APD. A combination of a half-wave plate (HWP), polarising beam cube ($\rm{bc}_{2}$) and quarter-wave plate (QWP) is used to ensure circular polarisation at the objective. A photodiode (Thorlabs DET10A) samples the laser at $\rm{bc}_{2}$ just before the fibre as the reference for the time-of-flight measurement. The inset shows the configuration used for characterisation of the fibre TM where the collecting fibre is replaced with a camera.}
\end{figure*}

\noindent{\bf Imaging procedure}.
For image acquisition the transmission matrix is used to calculate a series of binary holograms for display on the DMD. These holograms are designed to generate a single point in the far field of the fibre output facet which can be raster scanned across a scene. For the generation of a point, the corresponding row of the transmission matrix provides the complex weights for each respective input plane wave mode emanating from the DMD. This is equivalent to assuming that the inverse of the measured TM is equal to its conjugate transpose (i.e.\ the TM is unitary), which is a reasonable assumption for propagation through an optical fibre followed by free-space in which power loss is small within the NA of the system. The weighted sum of plane waves then gives a complex field which is to be displayed on the DMD for generation of the desired point. The complex field is converted into a binary hologram as by thresholding the argument. This method neglects the different amplitude modulations required for each DMD pixel. However, upon investigation, in our case incorporating amplitude modulation seemed to make little difference to the imaging fidelity and spot contrast, yet had a significant cost to total output power. Therefore we use phase only modulation.

Each row of the transmission matrix corresponds to an output spot position in a rectilinear grid. A subset of these positions will be used in a scan for imaging and their corresponding rows will be converted into binary holograms for the DMD as described above. Image acquisition is performed by first loading these holograms onto on-board memory of the DMD, and then cycling through them to scan the scene in the far field of the fibre. The DMD is triggered by the rising edge of the laser trigger output. The laser repetition rate is approximately 21\,kHz which translates to an image rate of 5Hz for a circular field of view with a 75-point diameter. For time-of-flight imaging we collect the reflected light in a 15 cm long large core secondary fibre which is coupled to an avalanche photodiode (Menlo Systems APD210). The output of the avalanche photodiode is recorded with a high-speed oscilloscope (Picoscope 6407) collecting samples at 2.5~GHz. The oscilloscope acquisition is triggered by the rising edge of a reference signal recorded by the secondary photodiode, located just before the objective. Once triggered, the oscilloscope saves the APD voltage of both reference and reflected signal to onboard memory. The recorded data are taken from one sample point before the trigger to a user defined number of sample points after. The number of samples collected is varied to increase data transfer efficiency when imaging objects nearer to the fibre. After recording a histogram, the following laser pulse triggers the DMD to load the next hologram. The data is then transferred to a computer for processing.\\

\noindent{\bf Time-of-flight processing}.
In order to generate a 3D image of the scene which has been scanned, two things must be determined from each point and therefore each recorded histogram. These are the depth, which is determined from the time to peak maximum, and the reflectivity, which is taken as proportional to the peak maximum value. Despite the high sampling rate, a 700\,ps pulse corresponds to just three or four sample points recorded above background by the APD. Hence, to more precisely determine the peak location, a sinc interpolation is performed on the data increasing the number of sample points by a factor of 32. The data then undergo temporal registering. This is performed by interpolating the trigger signal and finding the time bin where it first crosses a set threshold. All data points in the interpolated histogram before this threshold is passed are truncated for the recorded signal.

We synchronise every recorded point in the raster scan using its corresponding trigger signal. This has the effect of temporally registering all the recorded histograms such that they have the same ’zero’ with respect to the laser pulse from which timing can start. A universal time axis is then created for all raster scan points which is offset to account for the time the pulse spends in the fibres and the optical setup. The time axis is then squared and multiplied by each histogram to account for the inverse square reduction in reflected intensity. The data are then further truncated between user set limits which define the depth range over which the imaging is taking place. These limits need not be very strict, however, we note that when imaging poorly reflecting objects close to the fibre it is helpful to set an upper limit for the depth as the APD noise at longer distances can be over emphasised by the inverse square correction to the point of dominating the actual signal. The distance for each point in the raster scan is determined by finding the time corresponding to the maximum in the histogram and multiplying by half the speed of light. The reflectivity is simply the value of this maximum of the corrected histogram. The image produced is then thresholded by a proportion of the reflectivity maximum, typically 0.05 to 0.1, to reduce the effects of noise. Finally a field-of-view flattening step is performed where the depth is corrected to account for equidistant points from the fibre facet forming a sphere rather than a plane. This is performed as follows, 
\begin{equation}D_{\rm{flattened}}\left(r\right) = D\left(r\right) \cos\left(\theta\right)
\end{equation}
where $D=\frac{1}{2}c\delta t$ is the recorded depth, and $r$ is the radius in pixels measured from the centre of the field of view. $\theta$ is the polar angle and in practice is calculated through
\begin{equation}
  \theta = \tan^{-1}\left(\frac{r}{r_{\rm{max}}}\tan\left(\sin^{-1}\left(\rm{NA}\right)\right)\right),
\end{equation}
where $r_{\rm{max}}$ is the maximum radius of the field of view in pixels, and NA is the emission fibre numerical aperture which is assumed to be equivalent to the angle at this maximum radius. The equivalent equation in the main text, Eqn.\ 2, also includes a second term to account for angular differences in optical path length inside the fibre, which is derived below. We did not include this additional term in our data processing as it was negligible.
After processing, the data are then composited into frames of a video and replayed at the image acquisition rate for viewing.\\
\begin{figure}
    \centering
    \includegraphics[width=0.45\textwidth]{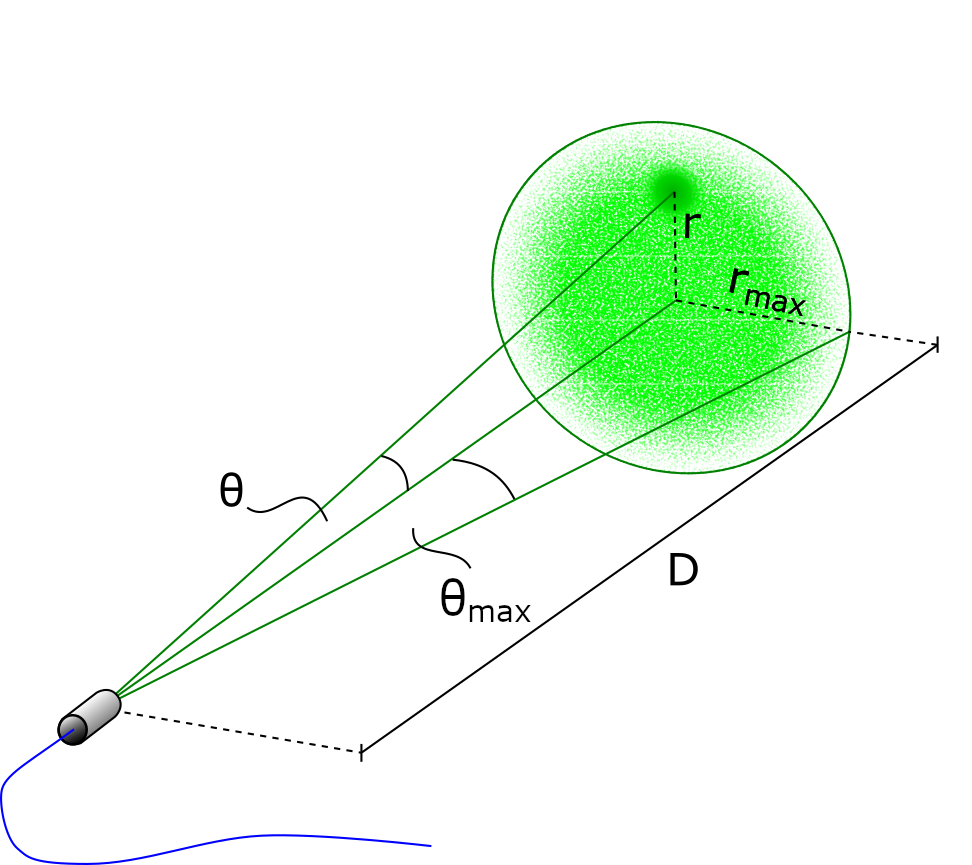}
    \caption{Illustration of the coordinate layout of the imaging system as used in the text. The maximum angle is given by $\theta_{\rm{max}}=\sin^{-1}\left(\rm{NA}/\rm{n}_{\rm{air}}\right)$.}
    \label{fig:coordinates}
\end{figure}

\noindent{\bf Calculation of depth from time-of-flight}. 
Pulses projected into the far-field of the fibre output facet travel a projection angle dependent optical path length (OPL) through the system. This is factored into the calculation of the depth of each pixel according to main text Eqn.~2, which is derived as follows. The round-trip flight time $\delta t$ is given by:
\begin{equation}\label{Eqn:opl}
    \delta t = \rm{OPL}/c = 2\left(\rm{OPL}_{\rm{mmf}} + \rm{OPL}_{\rm{fs}}\right)/c.
\end{equation}
Here $\rm{OPL}_{\rm{mmf}}$ is the optical path length of light propagating one way through the MMF, and  $\rm{OPL}_{\rm{fs}}$ is the optical path length of light propagating through the region of free-space in between the fibre output facet and the object. The factor of 2 on the right-hand-side of Eqn.~\ref{Eqn:opl} captures the double-pass through the system due to the round-trip of the pulse. The projection angle dependence of $\rm{OPL}_{\rm{mmf}}$ and $\rm{OPL}_{\rm{fs}}$ is given by:
\begin{equation}
    \rm{OPL}_{\rm{mmf}} = \frac{n_{\rm{c}}L}{\cos(\theta_{\rm{c}})},
\end{equation}
\begin{equation}
    \rm{OPL}_{\rm{fs}} = \frac{D}{\cos(\theta)},
\end{equation}
where $\theta$ is the projection angle of the pulse from the fibre facet, with respect to the optical axis of the the fibre, and $\theta_{\rm{c}}$ is the corresponding angle of propagation of the pulse within the fibre core. $\theta$ and $\theta_{\rm{c}}$ are related through Snell's law: $\sin(\theta) = n_{\rm{c}}\sin(\theta_{\rm{c}})$, and so:
\begin{equation}
    \cos(\theta_{\rm{c}}) = \left(1-\frac{\sin^2\theta}{n_{\rm{c}}^2}\right)^{\frac{1}{2}},
\end{equation}
where we have used the identity: $\cos(\sin^{-1}\left(x\right)) = \left(1-x^{2}\right)^{1/2}$. Therefore substituting the above into Eqn.~\ref{Eqn:opl} and rearranging for depth $D$ yields main text Eqn.~2:
\begin{equation}\label{Eq:flightDist}
   D\left(\delta t,\theta\right) = \frac{1}{2}c\delta t\cos\theta - n_{\rm{c}}L\cos\theta\left(1-\frac{\sin^2\theta}{n_{\rm{c}}^{2}}\right)^{-\frac{1}{2}}.
\end{equation}
It is straightforward to modify this expression to account for changes in ambient refractive index if imaging through fluid, and also to account for any differences in properties (such as length or core refractive index) between the illumination and collection fibres.\\

\noindent{\bf Calculation of minimum pulse duration}.
Main text Eqn.~1 provides a bound on the minimum pulse duration $\tau_{\rm{p}}$, as a function of fibre length $L$, to achieve {\it arbitrary} wavefront shaping at the output without temporal distortion. This can be derived as follows. To ensure pulse travelling at different speeds (thus different optical path lengths) through the fibre are still overlapping at the output, we must ensure that:
\begin{equation}\label{Eqn:tauopl}
    \tau_{\rm{p}} >> 2\frac{\Delta_{\rm{OPL}}}{c}.
\end{equation}
Here the factor of 2 on the right-hand-side signifies a double pass round-trip through the system, and $\Delta_{\rm{OPL}} = \rm{OPL}_{\rm{L}}-\rm{OPL}_{\rm{S}}$, where $\rm{OPL}_{\rm{L}}$ and $\rm{OPL}_{\rm{S}}$ are the longest and shortest possible optical path lengths respectively. The longest optical path length is given by light travelling at the critical angle of total internal reflection through the fibre. The angle of propagation within the core is governed by the numerical aperture of the fibre ($\rm{NA}$) and the refractive index of the core ($n_{\rm{c}}$), yielding:
\begin{equation}
\rm{OPL}_{\rm{L}} = \frac{L n_{\rm{c}}}{\cos\left(\sin^{-1}(\rm{NA}/n_{\rm{c}})\right)}.
\end{equation}
The shortest optical path length is given by light travelling parallel to the longitudinal axis of the fibre: $\rm{OPL}_{\rm{S}} = Ln_{\rm{c}}$. Therefore combining the above and substituting into Eqn.~\ref{Eqn:tauopl} gives:
\begin{equation}
    \tau_{\rm{p}} >> \frac{2Ln_{\rm{c}}}{c}\left[\left(1-\frac{\rm{NA}^2}{n_{\rm{c}}^2}\right)^{-\frac{1}{2}}-1\right],
\end{equation}
where we have used the identity $\cos(\sin^{-1}\left(x\right)) = \left(1-x^{2}\right)^{1/2}$. This can be further simplified under the assumption that $\rm{NA}^2/n_{\rm{c}}^2<<1$: we can Taylor expanding the term in the curved brackets and truncate powers of order $x^2$ and higher, yielding main text Eqn.~1:
\begin{equation}
    \tau_{\rm{p}} >> \frac{\rm{NA}^2L}{cn_{\rm{c}}}.\\
\end{equation}

\noindent{\bf Calculation of imaging resolution}.
The number of spatial modes supported by the MMF at a single polarisation is $M$: 
\begin{equation}
    M \sim \frac{V^2}{4} = \frac{\pi^{2}a^{2}\rm{NA}^2}{4\lambda^2},
\end{equation}
where $V$ is the fibre $V$-number. The area of the field-of-view at a distance $l$ from the output facet of the fibre, $A_{\rm{fov}}$, is:
\begin{equation}
    A_{\rm{fov}} = \pi r^{2} = \pi \frac{l^{2}\rm{NA}^2}{\left(1-\rm{NA}^2\right)},
\end{equation}
where $r$ is the radius of the circular field-of-view, and we have used the fact that $r = l\tan(\sin^{-1}\left(\rm{NA}\right))$ and the identity $\tan(\sin^{-1}\left(x\right)) = x\left(1-x^{2}\right)^{-1/2}$. Therefore the area of each spot in the far-field is given by $A_{\rm{spot}}$:
\begin{equation}
    A_{\rm{spot}} = \frac{A_{\rm{fov}}}{M}\sim\frac{4l^{2}\lambda^{2}}{\pi a^{2}\left(1-\rm{NA}^{2}\right)},
\end{equation}
and so the expected radius of each far-field spot is given by:
\begin{equation}
    r = \left(\frac{A_{\rm{spot}}}{\pi}\right)^{\frac{1}{2}} \sim\left(1-\rm{NA}^{2}\right)^{-\frac{1}{2}}\frac{2}{\pi}\frac{l\lambda}{a}.
\end{equation}
The above agrees well with Fraunhofer diffraction from a circular aperture of radius $a$, i.e. the Rayleigh criterion:
\begin{equation}
    r\sim\frac{1.22}{2}\frac{l\lambda}{a}.
\end{equation}
As this is linear in the distance $l$ the angular radius is expected to be constant with distance.\\

\noindent{\bf Quantification of imaging performance}\\
\noindent{\it Signal-to-noise ratio and depth accuracy measurements.} White screens were each imaged 100 times at five distances (40\,cm, 80\,cm, 120\,cm, 160\,cm, 200\,cm). Regardless of the screen distance, 71 data points were collected, one before the trigger and 70 after. Allowing for an offset of 5.7\,ns, during which time the pulse is in the optical setup and fibres, the 70 samples allow for imaging distances up to 334.5\,cm which generously covers the furthest screen distance.

\noindent{\it Signal-to-noise-ratio calculation}.
Each histogram from each point in the raster scan was individually standardised so that the minimum value was set to zero. The maximum value of each histogram was then recorded and 14 data points centred on this value were removed from the histogram. Since the pulse duration is $\sim$700\,ps, we typically observed a maximum of approximately 4 data points recorded, at 2.5\,GS/s, at levels significantly above background. Hence, removing the 14 points ensured that, provided the peak selected corresponded to the laser pulse, all that remained in the histogram was background recording.  The standard deviation of the remaining data points was then calculated and the ratio of peak to background standard deviation was calculated and averaged over the 100 frames. This resulted in an SNR map for each of the five measured distances.

\noindent{\it Depth error calculation}.
After processing the histograms as we did for normal imaging and averaging the depth map over the 100 frames, we standardised the data by subtracting the mean depth in each of the five maps. This was done to ensure errors in misalignment of the screens were not the reason for inaccurate depth readings. An ANOVA test showed no significant difference between the estimated depth accuracy at each of the screen distances so the five distances were averaged into a single depth map.
 
\noindent{\it Resolution and contrast methods and calculations}.
For estimation of the resolution and spot contrast a white screen placed at 20cm from the fibre facet was used. The points used in the raster scan were then displayed on the screen and the resulting pattern was recorded using the same camera used in the generation of the transmission matrix. For each of the images recorded in the raster scan an ideal Airy-disc was fit by optimising the width, x-location, y-location, and amplitude so as to minimise the least squares difference between the image and the ideal disc. To ensure no spurious fits skewed the data certain criteria were used as thresholds to remove poor fits. These included; if the fit centre was significantly different from the expected location of the spot, if the spot width was more than four times the expected width, and if the fit centre fell outside 1.1 times the fibre’s numerical aperture. Once this thresholding operation was completed, a resolution map was generated from the spot widths and a contrast map from the ratio of the intensity in the spot to the total intensity.

\noindent{\it Presentation of all 1D plots in Figure 3}.
The recorded data maps were then grouped by their angular position in the imaging field-of-view where the angle was measured as the deviation from the optical axis. For each bin the arithmetic mean and standard deviation were taken. Results are displayed in Fig.~3.\\

\noindent{\bf Transmission matrix guided computational imaging}.
 In Fig.~4 we use a more elaborate fibre calibration and reconstruction algorithm to create a depth map of higher fidelity. We term this method {\it transmission matrix guided computational imaging}. In this case we calibrate the fibre multiple times with different internal references, each forming a different reference speckle pattern in the far-field. These TMs are then combined to yield a single TM in which the blind spots have been removed and so the spot can be scanned to all output locations with good contrast. We next make a set of additional calibration measurements, by recording the intensity pattern at the output when each spot is projected. For each scan location, these measurements capture the contrast of the spot, and also the structure of the residual speckle pattern surrounding it. Once an image of a scene is recorded, we use our additional calibration measurements to reconstruct the depth image as follows:
 
 Rather than assuming that all power was focused into each spot location, we now incorporate knowledge of the actual projected speckle patterns and create a series of images as a function flight time (i.e. depth $D$). To achieve this, for each image we aim to solve the matrix-vector equation
 \begin{equation} \label{eq:ls}
 {\bf A\bf x} = {\bf y},
\end{equation}
 where ${\bf A}$ is a matrix that encapsulates the pre-measured projected speckle patterns (each measured speckle pattern is vectorised and forms a single row of ${\bf A}$), ${\bf x}$ is a column vector representing the image we aim to reconstruct in vectorised form, and ${\bf y}$ is a column vector of the measured intensity at a time corresponding to depth $D$, for each projected pattern. In order to accurately sample the projected speckle patterns, ${\bf A}$ typically has significantly more columns than rows, rendering Eqn.~\ref{eq:ls} undersampled. To overcome this we use a form of generalised Tikhonov regularisation to suppress noise. Each image is recovered by solving  the following optimisation problem
 \begin{equation}\label{Eq:Tik}
     \hat{{\bf x}} = \mathrm{arg min}\left|\left|{\bf Ax}{\bf-y}\right|\right|^2_2 + \left|\left|{\lambda {\bf x}}-{\bf A}^{\dagger}{\bf{y}}\right|\right|^2_2,
 \end{equation}
where $\hat{{\bf x}}$ is the final image and ${\bf x}$ is the decision variable. The first term on the right hand side constrains ${\bf x}$ to agree with the projected measurements as in Eqn.~\ref{eq:ls}. An approximate solution to Eqn.~\ref{eq:ls} is given by ${\bf x}' = {\bf A}^{\dagger}{\bf{y}}$, where $(\cdot)^{\dagger}$ denotes the conjugate transpose operation. ${\bf x}'$ is equivalent to a weighted sum of the projected speckle patterns, each weighted by the corresponding measurement held in vector ${\bf y}$. As in our case approximately 40\% of the projected power is focussed into each target spot, ${\bf x}'$ is typically a smooth function that peaks in the same regions at the real solution. Therefore the second term on the RHS constrains ${\bf x}$ to be close to ${\bf x}'$, which serves to suppress noise in the reconstruction. Here $\lambda$ is a tunable factor chosen to ensure proper scaling between ${\bf x}$ and ${\bf x}'$. As both terms on the RHS of Eqn.~$\ref{Eq:Tik}$ require minimisation of the square of the Euclidean norm, it is straight forward and fast to solve as a single matrix-vector equation using a standard least square solver, where the single matrix-vector equation becomes
 \begin{equation}
 \left[ \begin{array}{c} {\bf A} \\ \lambda{\bf I} \end{array} \right] {\bf x} = \left[ \begin{array}{c} {\bf y} \\ {\bf A}^{\dagger}{\bf y} \end{array} \right].
\end{equation}
This method works well for our data sets, however we note that a range of alternative regularisation methods are possible, incorporating a more rigorous consideration of anticipated noise levels. In addition, more sophisticated techniques related to compressed sensing can be employed, depending upon the level and form of prior information about the scene that is available.

Once the set of images at a series of depths have been calculated, a 3D surface image of the scene is obtained as follows: we loop through each lateral coordinate, at each point extracting the vector encapsulating the variation in intensity with depth. This vector is up-sampled using spline interpolation, and the depth at which the intensity peaks is located. This depth is assigned to the depth of the corresponding lateral pixel coordinate in the final 3D surface. If we expect the surface depth to vary smoothly with lateral coordinate (as is the case for the profile of the mannequin head shown in Figure 4), we also smooth each image with a Gaussian convolution kernel to further suppress noise.

We note that the signal-to-noise ratio (SNR) of single-pixel projective imaging methods (in which a scene is illuminated by a series of projected optical patterns and the level of back-scattered light captured with a single element detector) is optimised if all available power is focused into a single point, which is raster scanned. Therefore, our transmission matrix guided computational imaging approach relies on knowledge of the TM to attempt to focus all power into a single spot to optimise SNR, but then also accounts for any stray light forming a speckled background in the manner described above.

\end{document}